# Electronic and magnetic properties of orthorhombic iron selenide (FeSe)

S. W. Lovesey[1,2]

[1]ISIS Facility, STFC Oxfordshire OX11 0QX, UK

[2]Diamond Light Source Ltd, Oxfordshire OX11 0DE, UK

**ABSTRACT** Iron orbitals in orthorhombic iron selenide (FeSe) can produce charge-like multipoles that are polar (parity-odd). Orbitals in question include Fe(3d), Fe(4p) and p-type ligands that participate in transport properties and bonding. The polar multipoles may contribute weak, space-group forbidden Bragg spots to diffraction patterns collected with x-rays tuned in energy to a Fe atomic resonance (Templeton & Templeton scattering). Ordering of conventional, axial magnetic dipoles does not accompany the tetragonal-orthorhombic structural phase transition in FeSe, unlike other known iron-based superconductors. We initiate a new line of inquiry for this puzzling property of orthorhombic FeSe, using a hidden magnetic-order that belongs to the m'm'm' magnetic crystal-class. It is epitomized by the absence of ferromagnetism and axial magnetic dipoles, and the appearance of magnetic monopoles and magneto-electric quadrupoles. A similar magnetic order occurs in cuprate superconductors, YBCO & Hg1201, where it was unveiled with the Kerr effect and in Bragg diffraction patterns revealed by polarized neutrons.

## I. INTRODUCTION

Electronic, structural and magnetic properties of iron selenide, a tetragonal PbO-type Fe chalcogenide at room temperature, have attracted much attention following the discovery of iron-based superconductors. Superconductivity in β-FeSe occurs at ≈ 9 K in samples prepared with Se deficiency [1]. In common with the parent phases of iron pnictide superconductors the binary compound undergoes a tetragonal-orthorhombic transition, observed at ≈ 90 K [2], but the transition is not followed by long-range magnetic order [3-6]. A tetrad axis of rotational symmetry at Fe sites is lost in the structural transition that together with absence of time-reversal violation leads to a concept of nematic ordering. There is no consensus as to whether such order is driven by orbital, spin or lattice degrees of freedom, but evidence favouring an orbital mechanism has accumulated [6, 7]. An Ising-nematic transition features in models with quasi-local magnetic moments and frustrated Heisenberg interactions [8-12].

In this communication, we demonstrate that the currently accepted orthorhombic chemical structure adopted below ≈ 90 K possesses intriguing electronic properties visible in the Bragg diffraction of x-rays, although there is no published experimental data. The proposed experiments are a direct probe of Fe(3d) and p-type ligand orbitals that participate in transport properties and bonding (covalency). The information extracted from the experiments will not by itself answer the question as to the actual mechanism behind the tetragonal-orthorhombic transition, albeit orbital ordering or magnetic fluctuations that predict the same orthorhombic space-group symmetry, Cmma (#67), unlike arguments for mechanisms that drive the re-entrant tetragonal phase of $Ba_{1-x}Na_x Fe_2As_2$ [7].

The motif of electronic orbitals in the orthorhombic structure of FeSe manifests itself through polar (parity-odd) Fe multipoles that produce Bragg spots not indexed on the chemical structure. Similar, weak x-ray diffraction by axial (parity-even) multipoles in other compounds is known as Templeton & Templeton (T & T) scattering; for a review of conventional T & T scattering see Dmitrienko *et al* 2005 [13], for example. Polar electronic multipoles are allowed in an acentric environment. The low symmetry of sites occupied by Fe ions in orthorhombic FeSe possesses just three symmetry operations - diad axes of rotation symmetry along all cell edges and nothing more. The corresponding point group 222 ($D_2$) is consequently enantiomorphic (chiral). Multipoles in question are constructed from electronic states with different orbital angular momenta, e.g., expectation values using Fe(3d)-Fe(4p) and Fe(3d)-Se(4p) admixtures of polar tensor-operators. A successful x-ray diffraction experiment will confirm or not the space-group at a detailed level of enquiry, and yield unique information on Fe-Se hybridization.

Magnetism and superconductivity in iron pnictides appear to be bed-fellows, in that the two properties are contiguous in the phase diagram, quite unlike conventional BCS-type materials were a trace quantity of a magnetic impurities can destroy superconductivity. Magnetism in unconventional superconductors is thus an important field of investigation for the moment, because understanding the origin of the magnetism might shed light on the enigmatic mechanism for superconductivity [14-17]. The idea that magnetism in Fe-based superconductors might possess uncommon properties is prompted by the recent discovery that two cuprates, YBCO & Hg1201, possess ordered magnetic charge in the pseudo-gap phase [18, 19]. Returning to FeSe, it is well-established that conventional magnetic order, using axial magnetic dipoles, does not accompany the tetragonal-orthorhombic phase transition, which sets FeSe in a class of one with regard to known iron pnictide superconductors [3-6]. Our study, using magnetic symmetry, makes a case for hidden magnetic-order in orthorhombic FeSe that belongs to the m'm'm' magnetic crystal-class, within which axial magnetic dipoles are forbidden, and the Kerr effect and a fully compensated, antiferromagnetic motif of higher-order axial magnetic multipoles, illustrated in Figure 1, are allowed. Whence, axial magnetic quadrupoles and octupoles may contribute (1, 1, 0) and (1, 3, 0) Bragg spots, say, in diffraction patterns collected with neutrons or x-rays.

Magnetic charge, typified by a magnetic monopole, is notable by its absence in Maxwell's equations that unite electricity and magnetism. Artificially inserted in the equations, with symmetries of the electric and the magnetic field unchanged, magnetic charge is both time-odd and parity-odd like Dirac's magnetic monopole (a magneto-electric quantity). However, the magnetic charge in question is a property possessed by ions while Dirac's monopole - yet to be observed - is a fundamental unit of magnetic charge just like its sibling, the electron, is the fundamental unit of electric charge in Maxwell's equations.

The discovery of ordered magnetic charge in the pseudo-gap phase of cuprates was based on data collected by neutron diffraction [20, 21]. Bragg spots could be indexed on the chemical structure (space-group allowed reflections), and the intensities fit a ferro-type motif of magnetic charge in the form of magneto-electric quadrupoles. Exactly the same scenario is allowed for FeSe using the magnetic space-group Cm'm'a' (#67.507 in BNS setting) employed in the calculations using axial magnetism already mentioned (descriptions of magnetic space-groups, Belov–Neronova–Smirnova settings (BNS) and Wyckoff positions can be found at reference [22]). The predicted motif of magneto-electric quadrupoles in orthorhombic FeSe is shown in Figure 2, together with images of the quadrupoles and their response to inversion of either space or time. A central difference between the magnetic structures Cm'm'a' for FeSe and Cm'm'm' (#65.487) for Hg1201 is that, the former compound is allowed both axial magnetism and magnetic charge, whereas in the cuprate superconductor axial magnetism is forbidden by symmetry (time inversion and spatial inversion are conjugate symmetries at Cu sites in Hg1201, and a site symmetry that contains $\bar{1}'$, of which there are 18 types, forbids axial magnetism [23]). On the other hand, Cm'm'a' and Cm'm'm' both belong to the m'm'm' magnetic crystal-class, with the Kerr effect allowed whilst ferromagnetism is forbidden.

Properties of multipoles, and the magnetic space-group Cm'm'a' for FeSe, required to calculate unit-cell structure factors for Bragg diffraction are sketched in the following section. Thereafter, sections III & IV contain expressions for x-ray and neutron structure factors, and section V is given over to a survey of our principal conclusions.

## II. ELECTRONIC AND MAGNETIC PROPERTIES

Tetragonal iron selenide has the same basic structure as the iron arsenides. The space group is P4/nmm (#129) and Fe ions use sites 2a (3/4, 1/4, 0) with symmetry $\bar{4}$m2 ($D_{2d}$). The orthorhombic structure adopted below about 90 K is Cmma (#67), and Fe ions use sites 4a (1/4, 0, 0) with comparatively much lower symmetry designated by 222 (inversion, mirror and $\bar{4}$ symmetry elements are absent in an enantiomorphic point-group) [2, 3, 24]. But both tetragonal and orthorhombic structures are centro-symmetric, namely, crystal classes 4/mmm and mmm. Positions of Fe ions in Cmma are illustrated in Figures 1 & 2. A basis (x, y, z) for the low symmetry structure relative to the parent is obtained by a rotation of 45° about the c-axis. We use the basis {(1, 1, 0), (−1, 1, 0), (0, 0, 1)} and Miller indices for the orthorhombic structure are $h = H_o + K_o$, $k = -H_o + K_o$, $l = L_o$, where $H_o$, $K_o$, $L_o$ index reflections from the parent, P4/nmm.

Electronic degrees of freedom are encapsulated in multipoles that we denote $\langle O^K_Q \rangle$, with rank K and projections Q which obey $-K \leq Q \leq K$ [25]. Angular brackets signify an expectation value, or time average, of the enclosed spherical tensor-operator, and multipoles are properties of the electronic ground-state. A multipole is defined to have definite discrete symmetries, and $\sigma_\theta = \pm 1$ and $\sigma_\pi = \pm 1$ are its time and parity signatures, respectively. Our multipoles are derived from Hermitian operators, and the complex conjugate $\langle O^K_Q \rangle^* = (-1)^Q \langle O^K_{-Q} \rangle$, with $\langle O^K_Q \rangle = \langle O^K_Q \rangle' + i \langle O^K_Q \rangle''$ for real and imaginary parts. As we have already noted, the site symmetry 222 contains only diad axes of rotation symmetry. Requirements imposed by the diads are Q = 2p (p integer) together with $\langle O^K_Q \rangle = (-1)^K \langle O^K_{-Q} \rangle = (-1)^K \langle O^K_Q \rangle^*$.

Evidently, $\langle O^K_0 \rangle = 0$ for K odd, so dipoles (K = 1) are forbidden, and $\langle O^K_Q \rangle$ is purely real (imaginary) for K even (odd). Notably, a monopole $\langle O^0_0 \rangle$ is allowed in 222.

A unit-cell structure factor for Bragg diffraction is related to [26],

$$\Psi^K_Q = \sum_{\mathbf{d}} \exp(i\mathbf{d} \cdot \mathbf{k}) \langle O^K_Q \rangle_{\mathbf{d}}, \tag{1}$$

where the Bragg wavevector $\mathbf{k} = (h, k, l)$, and sites labelled $\mathbf{d}$ in a cell are occupied by Fe ions. C-centring in the orthorhombic structure is responsible for the selection rule $h + k = 2K_o$ even. A second selection rule arises from a factor $[1 + (-1)^h \sigma_\pi]$ in $\Psi^K_Q$, namely, $h$ odd (even) for parity-odd (parity-even) multipoles. Reflections $h + k$ = even with $h$ odd are space-group forbidden, and they can exist because Fe ions occupy acentric sites where $\sigma_\pi = -1$ is allowed. A mirror operation on a cell edge that relates two environments in a cell is actually the origin of $\sigma_\pi$ in $\Psi^K_Q$. Application of such a mirror operation to $\langle O^K_Q \rangle$ amounts to the inversion operation, and a factor $\sigma_\pi$, by virtue of the invariance of $\langle O^K_Q \rangle$ with respect to a two-fold, diad, rotation. Later we provide a unit-cell structure factor for resonant x-ray Bragg diffraction by polar multipoles. Thomson and nuclear scattering from Se will occur for $h + k$ = even, $l$ different from zero, while it is forbidden at $(h, k, 0)$ with $h, k$ odd.

Let us consider a hidden magnetic-order in the orthorhombic phase that is consistent with the known absence of axial magnetic dipoles [3-6]. Of the various compatible magnetic structures only one, Cm'm'a', forbids both magnetic dipole moments and diffraction at space-group allowed reflections. As for magneto-electric multipoles, with $\sigma_\theta = -1$ & $\sigma_\pi = -1$, a magnetic monopole (charge) is allowed but anapoles (toroidal dipoles) are forbidden. Sites used by Fe ions possess the same symmetry, 222, in both the parent Cmma and the magnetic space-groups Cm'm'a'. However, translation operations in Cmma and Cm'm'a' are different, and a mirror operation in Cmma is an anti-mirror in Cm'm'a'. Thus, the factor $[1 + (-1)^h \sigma_\pi]$ in $\Psi^K_Q$(Cmma) becomes $[1 + (-1)^h \sigma_\theta \sigma_\pi]$ in the result for $\Psi^K_Q$(Cm'm'a') that appears in (2). (Strictly speaking, paramagnetic site symmetry 2221' hosts indistinguishable site symmetries 222 and 2'2'2', and the magnetism allowed by Cm'm'a' selects 222.) While Cmma forbids axial magnetic dipoles it allows bulk magnetism and diffraction at space-group reflections. It is rejected as a candidate for magnetic order in orthorhombic FeSe on the grounds that the allowed, axial magnetism would have already been observed.

For Fe ions in FeSe described by the magnetic space-group Cm'm'a' we find the exact result,

$$\Psi^K_Q = \langle O^K_Q \rangle \exp(i\pi h/2) [1 + (-1)^{h+k}][1 + (-1)^h \sigma_\theta \sigma_\pi]. \tag{2}$$

Evidently, conditions on $\langle O^K_Q \rangle$ derived from the site symmetry 222 apply also to $\Psi^K_Q$, so dipoles are forbidden, i.e., $\Psi^1_Q = 0$ independent of the actual signatures for the discrete symmetries of $\langle O^K_Q \rangle$. Violation of time reversal symmetry means $\sigma_\theta = -1$. We suggest that the lack of experimental evidence for time-reversal violation in orthorhombic FeSe is actually a consequence of inappropriate experiments to date, and go on to suggest scattering experiments that will test the current proposal based on the magnetic space-group Cm'm'a'. Conventional, axial magnetism is consistent with $\sigma_\theta = -1$ & $\sigma_\pi = +1$, and it contributes to Bragg spots indexed by Miller index $h$ odd. This type of magnetism has not been observed in bulk magnetization measurements, because $\Psi^K_Q = 0$ for the ferromagnetic reflection $h = k = l = 0$, while the magnetic dipole moment $\langle \mathbf{L} + 2\mathbf{S} \rangle = 0$ due to Fe site symmetry. Magnetic charge is defined by $\sigma_\theta = -1$ & $\sigma_\pi = -1$, and it contributes to space-group allowed reflections indexed by Miller index $h$ even. Spin, $\mathbf{S}$, and the electric dipole, $\mathbf{n}$, can be used to represent magneto-electric multipoles [23, 25, 26]. In which case, $\langle \mathbf{S} \cdot \mathbf{n} \rangle$ and $\langle \mathbf{S} \times \mathbf{n} \rangle$ are a magnetic monopole and an anapole, respectively, and $\langle \mathbf{S} \times \mathbf{n} \rangle = 0$ due to Fe site symmetry. Corresponding quadrupoles are sketched in Figure 2.

### III. RESONANT X-RAY DIFFRACTION

The scattering amplitude for resonant x-ray Bragg diffraction is derived from quantum-electrodynamics [25, 26]. In the derivation, the QED amplitude is developed in the small quantity $E/mc^2$ where $E$ is the primary energy and $mc^2 = 0.511$ MeV. At the second level of smallness in this quantity the amplitude contains resonant processes that may dominate all other contributions to the amplitude should $E$ match an atomic resonance with an energy $\Delta$. Assuming also that virtual intermediate states are spherically symmetric, to a good approximation, the scattering amplitude $\approx F_{\mu'\nu}/(E - \Delta + i\Gamma/2)$ in the region of the resonance, where $\Gamma$ is the total width of the resonance. The numerator $F_{\mu'\nu}$ is a unit-cell structure factor for Bragg diffraction in the scattering channel with primary (secondary) polarization $\nu$ ($\mu'$). States of photon polarization are defined in Figure 3, and our unit-cell structure factors include dependence on the rotation of the crystal through an angle $\psi$ around the Bragg wavevector in a so-called azimuthal-angle scan. Intensity of a Bragg spot is proportional to $|F_{\mu'\nu}/(E - \Delta + i\Gamma/2)|^2$.

We explore consequences of the result (2) for charge-like multipoles ($\sigma_\theta = +1$) that are also polar ($\sigma_\pi = -1$). Multipoles with these discrete symmetries, denoted by $\langle U^K_Q \rangle$, are visible in resonant x-ray Bragg diffraction enhanced by a parity-odd absorption event [26, 27]. An E1-E2 event that gives access to polar multipoles with ranks $K = 1, 2, 3$ can exist in the vicinity of the Fe K-edge at an energy 7.112 keV. However, dipoles are forbidden in the orthorhombic structure, and E1-E2 unit-cell structure factors are a sum of quadrupoles ($K = 2$) and octupoles ($K = 3$). We consider reflections indexed by ($h, h, l$) and ($h, -h, l$) with $h$ odd. Quadrupoles (octupoles) are purely real (imaginary), which means $\Psi^2_Q$ is purely imaginary and $\Psi^3_Q$ is purely real. There are three unknowns in unit-cell structure factors, namely, $\Psi^2_0 \propto i\langle U^2_0 \rangle$, $\Psi^2_{+2} \propto i\langle U^2_{+2} \rangle'$ and $\Psi^3_{+2} \propto \langle U^3_{+2} \rangle''$.

Let $\gamma$ be the angle enclosed by ($h, h, l$) and the c-axis,

$$\cos(\gamma) = (la/c)/[h^2(1 + (a/b)^2) + (la/c)^2]^{1/2},$$

with $\gamma = 90°$ for Miller index $l = 0$ (cell lengths for the orthorhombic structure $a \approx 5.30781$ Å, $b \approx 5.33423$ Å and $c \approx 5.48600$ Å [2]). The unit-cell structure factor for the $\sigma'\sigma$ channel has a simple dependence on Bragg angle given by $\sin(\theta)$, whereas there is no simple dependence in the rotated channel $\pi'\sigma$. Using (2) we find,

$$F_{\sigma'\sigma}(E1\text{-}E2) = -i \sin(\theta) \sqrt{(1/15)} \, [-(\sqrt{3}/2) \sin^2(\gamma) \sin(2\psi) \, \Psi^2_0$$

$$+ \sqrt{2} \cos(\gamma) \cos(2\psi) \, \Psi^2_{+2}$$

$$+ i \cos(\gamma) \{[2 - 3\sin^2(\gamma)] \cos(2\psi) + 3\sin^2(\gamma)\} \, \Psi^3_{+2}], \quad (3)$$

for the unrotated $\sigma'\sigma$ channel. In the derivation of (3) we have neglected the small difference in cell lengths a and b. The corresponding expression for $F_{\pi'\sigma}(E1\text{-}E2)$ is lengthy, but it is also a sum of $\Psi^2_0$, $\Psi^2_{+2}$, $\Psi^3_{+2}$ [27].

The unit-cell structure factor (3) possesses uncommon features. It is valid for Bragg spots $(h, h, l)$ and $(-h, h, -l)$, and after a change in sign to $\Psi^2_{+2}$ and $\Psi^3_{+2}$ it is valid for $(h, -h, l)$ and $(-h, -h, -l)$. Additionally, $F_{\sigma'\sigma}(E1\text{-}E2)$ is the same at $(h, h, l)$ and $(-h, h, -l)$ because the two expressions are related by a simultaneous change in sign to $\Psi^K_{+2}$ and a shift $\gamma \to \gamma - \pi$. Exactly the same changes apply to $F_{\pi'\sigma}(E1\text{-}E2)$. Different intensity in Bragg spots $(h, h, l)$ and $(-h, -h, -l)$ violates Friedel's Law. Reflections $(h, -h, l)$ and $(h, h, l)$ in $\sigma'\sigma$ and $\pi'\sigma$ are a Bijvoet pair if a necessary number of multipoles is non-zero ($\Psi^2_0$ and one $\Psi^K_{+2}$ different from zero). Note that coefficients of $\Psi^K_{+2}$ in $F_{\sigma'\sigma}(E1\text{-}E2)$ are even functions of the azimuthal angle, while the coefficient of $\Psi^2_0$ is an odd function of $\psi$. The reverse behaviour is true in the $\pi'\sigma$ channel. Structure factors for rotated polarization obey $F_{\sigma'\pi}(E1\text{-}E2;\theta) = -F_{\pi'\sigma}(E1\text{-}E2;-\theta)$. Inspection of (3) shows that $F_{\sigma'\sigma}(E1\text{-}E2)$ is independent of $\Psi^K_{+2}$ for Miller index $l = 0$ but the same is not true of $F_{\pi'\sigma}(E1\text{-}E2)$. We find,

$$F_{\pi'\sigma}(E1\text{-}E2) = -(i/8\sqrt{5}) \, [(5\cos(2\theta) + 1)\cos(2\psi) + 2\cos^2(\theta)] \, \Psi^2_0$$

$$- (1/\sqrt{15}) \sin(2\theta) \sin(\psi) \, [i\sqrt{2} \, \Psi^2_{+2} + \Psi^3_{+2}], \quad (4)$$

at $(h, h, 0)$. (Thomson diffraction by Se ions is forbidden at $(h, k, 0)$ with $k$ odd.) Signs of $\Psi^2_{+2}$ and $\Psi^3_{+2}$ in (4) are reversed at $(h, -h, 0)$. Observation of Bragg spots $(h, h, 0)$ and $(h, -h, 0)$ can provide a measure of $\Psi^2_0$ relative to the combination of $\Psi^K_{+2}$ that appears in (4). The same result can be achieved by use of rotated polarizations $\pi'\sigma$ and $\sigma'\pi$. The experimental information on multipoles in question can be used to test *ab initio* calculations of electronic structure.

Additional symmetry in the tetragonal structure, adopted above the structural phase transition at $\approx 90$ K, restores Friedel's Law and eliminates a Bijvoet pair. Specifically, Q and parity are linked by $\overline{4}$, and Q = 2p with p odd for $\sigma_\pi = -1$. Space-group forbidden reflections for polar multipoles occur for $H_o + K_o$ odd. Reflections of the type $(H_o, 0, L_o)$ with $H_o$ odd are equivalent to those considered above for the orthorhombic structure. One finds $|F_{\sigma'\sigma}(E1\text{-}E2)|$ and $|F_{\pi'\sigma}(E1\text{-}E2)|$ are identical for $(H_o, 0, L_o)$ and $(-H_o, 0, -L_o)$. $|F_{\sigma'\sigma}(E1\text{-}E2)|$ is the same for $(H_o, 0, L_o)$ and $(-H_o, 0, L_o)$ but $|F_{\pi'\sigma}(E1\text{-}E2)|$ is different at the two reflections. While $F_{\sigma'\sigma}(E1\text{-}E2)$ is a function of $\cos(2\psi)$ the structure factor for rotated polarization is a sum of terms containing sin(ψ) and sin(2ψ). The odd harmonic of the azimuthal angle in $F_{\pi'\sigma}$(E1-E2) makes the difference between ($H_o$, 0, $L_o$) and (−$H_o$, 0, $L_o$).

## IV. MAGNETIC DIFFRACTION

We explore a plausible model of hidden magnetic-order described in §§ I, II, namely, magnetic space-group Cm'm'a' for which (2) is the correct electronic structure factor. Properties of the magnetic order are unavailable to bulk measurements. Fortunately, both neutron and x-ray Bragg diffraction experiments can unveil the magnetism of Fe ions in Cm'm'a'. We start with neutron Bragg diffraction, for which an introduction is likely not called for [23].

Parity-even multipoles $\langle T^K_Q \rangle$ contribute to space-group forbidden Bragg spots ($h$, $k$, $l$) with $h$ and $k$ odd integers. Expressions we provide for the magnetic neutron diffraction amplitude $\langle \mathbf{Q}_\perp \rangle$ neglect multipoles with a rank larger than 3, on the grounds that they likely contribute small corrections to multipoles of lower rank. It is convenient to separate $\Psi^K_Q$ into even and odd functions of Q and use,

$$A^2_Q = 4 \exp(i\pi h/2) \langle T^2_Q \rangle' \text{ and } B^3_2 = i4 \exp(i\pi h/2) \langle T^3_{+2} \rangle'', \tag{5}$$

derived from (2). Intensity of a Bragg spot is $|\langle \mathbf{Q}_\perp \rangle|^2$ with,

$$\langle \mathbf{Q}_{\perp,x} \rangle \approx \kappa_y \kappa_z \sqrt{3} \, [A^2_2 + \sqrt{(3/2)}A^2_0 + i \, (\sqrt{35}/2\sqrt{2}) \, B^3_2 \, (3\kappa_x^2 - 1)],$$

$$\langle \mathbf{Q}_{\perp,y} \rangle \approx \kappa_x \kappa_z \sqrt{3} \, [A^2_2 - \sqrt{(3/2)}A^2_0 + i \, (\sqrt{35}/2\sqrt{2}) \, B^3_2 \, (3\kappa_y^2 - 1)],$$

$$\langle \mathbf{Q}_{\perp,z} \rangle \approx \kappa_x \kappa_y \sqrt{3} \, [-2A^2_2 + i \, (\sqrt{35}/2\sqrt{2}) \, B^3_2 \, (3\kappa_z^2 - 1)]. \tag{6}$$

Here, a unit vector $\boldsymbol{\kappa} = \mathbf{k}/k$ and $\kappa_x \propto h$, $\kappa_y \propto k$, $\kappa_z \propto l$. These expressions are purely imaginary, and the corresponding intensity is in quadrature with intensity due to Se nuclei. For Bragg spots ($h$, $k$, 0) only $\langle \mathbf{Q}_{\perp,z} \rangle$ can be different from zero, and nuclear scattering by Se is forbidden.

By way of orientation to multipoles that could form a magnetic order-parameter we introduce a simple model of the ferrous ion (3d$^6$). A single electron outside a half-filled shell is likely to be strongly influenced by the lattice and hybridization with lattice vibration expected, as in FeF$_2$, for example. For the moment we make use of a high-spin configuration $^5$D for which the total angular momentum J = 4. A simple candidate wave-function that complies with symmetry 222 is,

$$|g\rangle = \cos(\beta) \, (1/\sqrt{2})[|2\rangle + |-2\rangle] - i \sin(\beta) \, |0\rangle, \tag{7}$$

where |M⟩ is short-hand for |J, M⟩. The mixing angle β is unknown. Multipoles of even rank are zero for any state that contains unique atomic quantum numbers, i.e., a wave-function derived from a manifold of states as in (7) [23]. Thus, the quadrupole $A^2_Q = 0$ for our current simple model. The octupole in $B^3_2$ is found to be,

$$\langle T^3_{+2} \rangle = i \, (5/49) \, \sqrt{(3/14)} \sin(2\beta) \, [-\langle j_2(k) \rangle + (4/3) \langle j_4(k) \rangle]. \tag{8}$$

Radial integrals $\langle j_2(k) \rangle$ and $\langle j_4(k) \rangle$ vanish in the forward direction k = 0 [28]. A quadrupole is proportional to $\langle j_2(k) \rangle$ and it can have a significant impact on neutron intensity as a function of k, as witnessed by a detailed examination of neutron diffraction by an iridate, $Sr_2IrO_4$ [29].

For a Bragg spot indexed by (h, k, 0) the result (8) provides the guide,

$$\langle \mathbf{Q}_{\perp,z} \rangle \approx - \kappa_x \kappa_y \, 0.685 \sin(2\beta) \, [\langle j_2(k) \rangle - (4/3) \langle j_4(k) \rangle], \qquad (9)$$

with $\langle \mathbf{Q}_{\perp,x} \rangle = \langle \mathbf{Q}_{\perp,y} \rangle = 0$. Using expressions in [28] for the radial integrals $\langle j_2(k) \rangle$ and $\langle j_4(k) \rangle$, for the ion $Fe^{2+}$, we find that (9) as a function of $k = (4\pi/\lambda) \sin(\theta)$ possesses a maximum value at $\sin(\theta)/\lambda \approx 0.32$ Å$^{-1}$ which is achieved at the (1, 3, 0) Bragg spot, to a good approximation.

Parity-even magnetic multipoles are responsible for Bragg spots in resonant x-ray diffraction indexed by *h* and *k* odd, and correspond to magnetic T & T scattering. For the parity-even E2-E2 absorption event at the K-edge multipoles relate to orbital properties of d-like valence electrons, and there is no information about spin degrees of freedom [30]. Magnetic parity-even multipoles have K odd. This condition and the triangle rule for E2-E2 imposes the values K = 1, 3 and site symmetry 222 ultimately restricts the rank to K = 3 with projections Q = ±2. Parity-even multipoles in resonant x-ray diffraction are denoted $\langle \check{T}^K_Q \rangle$, and they are purely imaginary in the present case with $\langle \check{T}^K_Q \rangle = i \langle \check{T}^K_Q \rangle''$. (In references [25, 26] multipoles in question are denoted by T instead of $\check{T}$, and here T is reserved for neutron diffraction.)

From (2), $\Psi^3{}_{+2} = i4 \exp(i\pi h/2) \langle \check{T}^3{}_{+2} \rangle''$ is purely real for *h* odd and,

$$F_{\sigma'\sigma}(E2\text{-}E2) = (\sqrt{3}/2) \sin(\gamma) \sin(2\theta) \sin(\psi) [\cos^2(\psi) - \cos^2(\gamma) \sin^2(\psi)] \, \Psi^3{}_{+2}, \qquad (10)$$

for the unrotated channel of polarization. Expression (10) is for (h, h, l) and (−h, h, −l), while for (h, −h, l) and (−h, −h, −l) there is nothing more than an overall change of sign. One finds $\theta \approx 45°$ for the reflection (3, 3, 0), based on E = 7.112 keV for the energy of the K-edge. The corresponding structure factor for the rotated channel of polarization reduces to $F_{\pi'\sigma}(E2\text{-}E2) \propto \sin^3(\psi) \Psi^3{}_{+2}$, to a good approximation.

Using the candidate wave-function (7),

$$\langle \check{T}^3{}_{+2} \rangle = - (i/28)\sqrt{(3/5)} \sin(2\beta). \qquad (11)$$

This gives a maximum value for $\langle \check{T}^3{}_{+2} \rangle''$ of 0.028.

We close the section with results for neutron diffraction by magneto-electric quadrupoles that reflect a state of ordered magnetic charge. Bragg spots are indexed on the chemical structure and both nuclear and magnetic scattering are allowed. Polarization analysis can be used to isolate the magnetic contribution [20]. Primary and secondary polarizations are **P** and **P'**, a fraction $(1 - \mathbf{P} \cdot \mathbf{P'})/2$ of neutrons change (flip) the neutron spin orientation, and $(1 - \mathbf{P} \cdot \mathbf{P'})/2 \propto \{(1/2) (1 + P^2) |\langle \mathbf{Q}_\perp \rangle|^2 - |\mathbf{P} \cdot \langle \mathbf{Q}_\perp \rangle|^2\}$ for a collinear magnetic motif. We provide results for an intermediate amplitude that appears in $\langle \mathbf{Q}_\perp \rangle^{(-)} = \boldsymbol{\kappa} \times (\langle \mathbf{Q} \rangle^{(-)} \times \boldsymbol{\kappa})$, where the superscript denotes that diffraction is by magneto-electric multipoles with $\sigma_\pi = -1$ (use of the same letter for projections, Q, and the intermediate scattering amplitude, **Q**, is an unfortunate outcome of the history of neutron scattering, and it should not be misleading in the present context). By limiting attention to the lowest-order multipoles, namely, quadrupoles $\langle H^2_Q \rangle$, we arrive at the simple expressions,

$$\langle \mathbf{Q}_x \rangle^{(-)} \approx \kappa_x\, C\, [\langle H^2_0 \rangle - \sqrt{6}\, \langle H^2_{+2} \rangle'],\ \langle \mathbf{Q}_y \rangle^{(-)} \approx \kappa_y\, C\, [\langle H^2_0 \rangle + \sqrt{6}\, \langle H^2_{+2} \rangle'], \quad (12)$$

$$\langle \mathbf{Q}_z \rangle^{(-)} \approx -2\, \kappa_z\, C\, \langle H^2_0 \rangle,$$

with $C = -2i\, \sqrt{(6/5)}\, (-1)^n$ and $h = 2n$. Magneto-electric quadrupoles in (12) are depicted in Figure 2, and they can be expressed as $\langle H^2_0 \rangle \propto \langle 3 S_z\, n_z - \mathbf{S} \cdot \mathbf{n} \rangle$ and $\langle H^2_{+2} \rangle' \propto \langle S_x\, n_x - S_y\, n_y \rangle$, where **S** and **n** are spin and electric dipole operators. Magnetic and nuclear scattering amplitudes differ in phase by 90°.

The radial integral included in $\langle H^2_Q \rangle$ is an expectation value of a spherical Bessel function of order 1, and it is a linear function of $k = (4\pi/\lambda) \sin(\theta)$ for small k. The radial integral for divalent Fe is discussed in a paper that reports calculations of diffraction amplitudes for the iron chalcogenide $BaFe_2Se_3$ [31]. Using atomic states $Fe(3d^6)$ and $Fe(4p^1)$ in an illustration, the Bragg spot (0, 2, 0) is within the first maximum of the corresponding radial integral. A better calculation of the radial integral will include an Fe(3d)-Se(4p) admixture, but this is unlikely to significantly change the dependence on k set by the spherical Bessel function of order 1.

## V. CONCLUSIONS

In summary, we have demonstrated that x-ray Bragg diffraction enhanced by a parity-odd Fe absorption event has the potential to provide unique information about the electronic properties of FeSe, while testing the established chemical structure Cmma (#67). Reflections (*h*, *h*, 0) with Miller index *h* odd in the orthorhombic structure are an attractive option, because Thomson diffraction by Se ions is forbidden. Iron polar multipoles exist because hybridization of valence states that differ in orbital angular momentum by an odd integer is allowed in an acentric environment. Such states may include Fe(3d)-Fe(4p) and Fe(3d)-Se(4p) overlap. Bragg intensities for the orthorhombic phase that exists below ≈ 90 K should display uncommon relations, due to the violation of Friedel's Law and formation of Bijvoet pairs.

A plausible candidate for hidden magnetic-order in the orthorhombic phase, Cm'm'a' [22], can be tested by the Kerr effect [33], and neutron diffraction and resonant x-ray diffraction. The corresponding magnetic crystal-class, m'm'm', forbids ferromagnetism, and the axial magnetic-order is a fully compensated, antiferromagnetic motif depicted in Figure 1. The candidate allows both axial (parity-even) and polar (parity-odd) magnetic multipoles, but dipoles of either type are forbidden. Absence of axial magnetic dipoles in the orthorhombic phase of FeSe is firmly established by experiments [3-6], which supports our use of Cm'm'a'. Moreover, a magnetic Bragg spot indexed (1, 1, 0), for example, is allowed by axial multipoles in our candidate while the (1, 0, 0) spot is forbidden, in accord with neutron diffraction data [32]. The hidden magnetic-order allows a Fe magnetic monopole that may contribute to x-ray diffraction enhanced by an E1-M1 event [25]. There is an interesting link between magnetic properties of our candidate for FeSe and magnetic order in high-$T_c$ superconductors, which has been established with the Kerr effect and by neutron Bragg diffraction [20, 21, 33]. For, the proposed ordered magnetic charge in orthorhombic FeSe, allowed by Cm'm'a' and depicted in Figure 2, is the same ferro-type order of polar magnetic quadrupoles recently discovered in the pseudo-gap phase of two cuprate superconductors, YBCO & Hg1201 [18, 19].

We close with a few remarks about calculating magnetic multipoles. Polar multipoles have been shown to be a product of Stone's model of electrons with locked spin and orbital degrees of freedom [34]. In the present investigation, an orientation to axial magnetic multipoles in the candidate magnetic structure is derived from a simple Fe wavefunction. Multipoles of interest for neutron and x-ray diffraction have been extracted from *ab initio* simulations of electronic structures derived recently for other compounds. These include hidden magnetic-order in URu$_2$Si$_2$ [35], and magnetic monopoles in lithium orthophosphates [36].

**Acknowledgements** I profited from assistance and diligent scrutiny from tutors at the ISIS Facility; Dr K S Knight with structural properties of FeSe, and Dr D D Khalyavin on every aspect of the proposed hidden magnetic-order and, also, with the preparation of Figures 1 & 2. Dr R Johnson (ISIS) commented on an early draft of the paper. Figure 3 was prepared by Professor E Balcar. Dr A Coldea, Professor S Margadonna, Professor S P Collins, Dr M D Watson and Professor P G Radaelli contributed with useful discussions and correspondence.

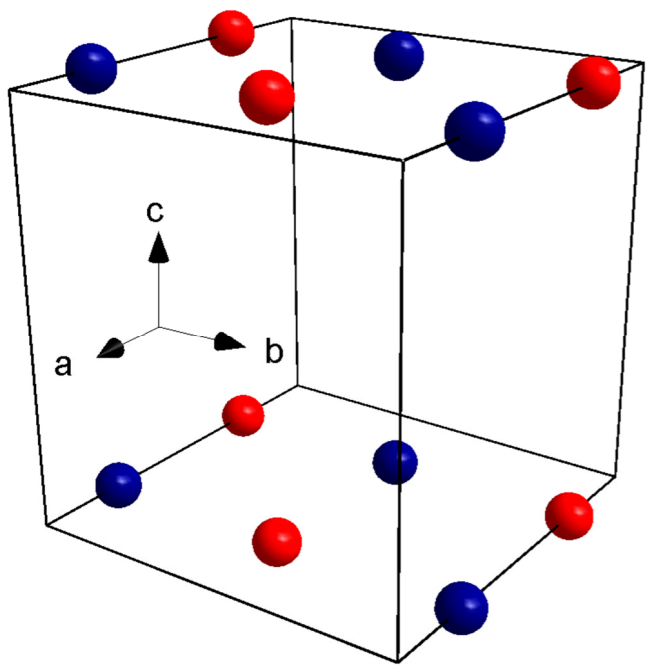

**Figure** 1 Depiction of a fully compensated, antiferromagnetic motif of Fe axial multipoles in orthorhombic FeSe derived from the magnetic space-group Cm'm'a'. Red and blue symbols have opposite time signatures. Se ions are not shown for the sake of clarity in Figures 1 & 2.

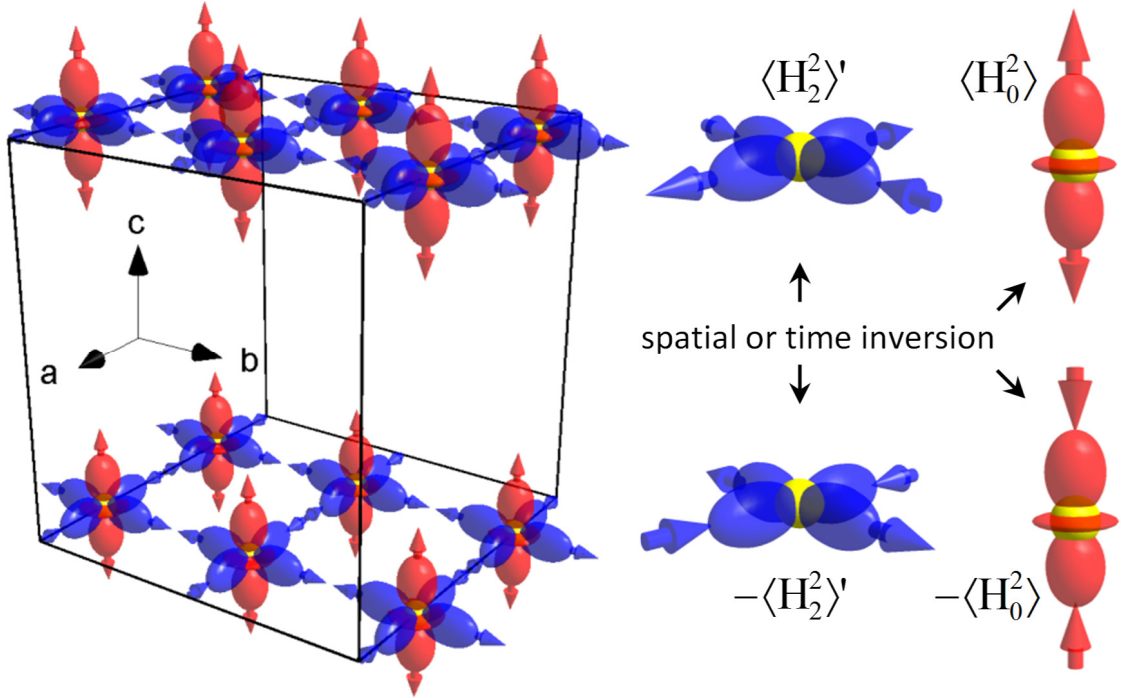

**Figure** 2 Ferro-type ordering of Fe polar magnetic multipoles in orthorhombic FeSe derived from the magnetic space-group Cm'm'a'. Arrows indicate spin directions in magneto-electric quadrupoles $\langle H^2_0 \rangle \propto \langle 3S_z n_z - \mathbf{S} \cdot \mathbf{n} \rangle$ and $\langle H^2_{+2} \rangle' \propto \langle S_x n_x - S_y n_y \rangle$ that occur in the amplitudes for neutron diffraction (12), together with their response to spatial or time inversion. Here, $\mathbf{S}$ and $\mathbf{n}$ are spin and electric dipole operators, respectively, and the basis (x, y, z) coincides with the orthorhombic cell edges.

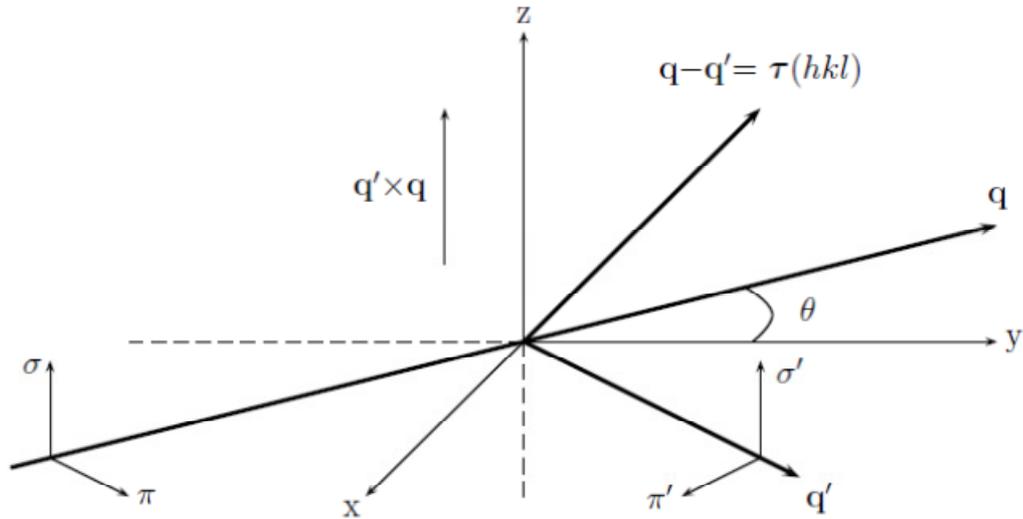

**Figure** 3 Cartesian coordinate system (x, y, z) adopted for resonant Bragg diffraction of x-rays and the relation to states of polarization, σ and π, in the primary (unprimed) and secondary (primed) beams. In the nominal setting of the crystal the system (x, y, z) coincides with basis vectors {(1, −1, 0), (1, 1, 0), (0, 0, 1)}, which are also given the same Cartesian labels. The beam is deflected through and angle 2θ, and **q** and **q'** are primary and secondary wavevectors. At the origin of an azimuthal scan (ψ = 0) a reciprocal-lattice vector (h, k, 0) is contained in the plane of scattering.